\documentclass[3p,,preprint,12pt]{elsarticle}
\makeatletter\if@twocolumn\PassOptionsToPackage{switch}{lineno}\else\fi\makeatother

\usepackage{tabulary,xcolor}
\usepackage{amsfonts,amsmath,amssymb}
\usepackage[T1]{fontenc}
\makeatletter
\let\save@ps@pprintTitle\ps@pprintTitle
\def\ps@pprintTitle{\save@ps@pprintTitle\gdef\@oddfoot{\footnotesize\itshape \null\hfill\today}}
\def\hlinewd#1{%
  \noalign{\ifnum0=`}\fi\hrule \@height #1%
  \futurelet\reserved@a\@xhline}

\AtBeginDocument{\ifNAT@numbers \biboptions{sort&compress}\fi}
\makeatother

\usepackage{ifluatex}
\ifluatex
\usepackage{fontspec}
\defaultfontfeatures{Ligatures=TeX}
\usepackage[]{unicode-math}
\unimathsetup{math-style=TeX}
\else 
\usepackage[utf8]{inputenc}
\fi 
\ifluatex\else\usepackage{stmaryrd}\fi

\usepackage{url,multirow,morefloats,floatflt,cancel,tfrupee}
\makeatletter

\AtBeginDocument{\@ifpackageloaded{textcomp}{}{\usepackage{textcomp}}}
\makeatother
\usepackage{colortbl}
\usepackage{xcolor}
\usepackage{pifont}
\usepackage[nointegrals]{wasysym}
\makeatletter

\def\mcWidth#1{\csname TY@F#1\endcsname+\tabcolsep}

\def\cAlignHack{\rightskip\@flushglue\leftskip\@flushglue\parindent\z@\parfillskip\z@skip}
\def\rAlignHack{\rightskip\z@skip\leftskip\@flushglue \parindent\z@\parfillskip\z@skip}

\@ifundefined{etal}{}{}

\usepackage{ifxetex}
\ifxetex\else\if@twocolumn\@ifpackageloaded{stfloats}{}{\usepackage{dblfloatfix}}\fi\fi

\AtBeginDocument{
\expandafter\ifx\csname eqalign\endcsname\relax
\def\eqalign#1{\null\vcenter{\def\\{\cr}\openup\jot\m@th
  \ialign{\strut$\displaystyle{##}$\hfil&$\displaystyle{{}##}$\hfil
      \crcr#1\crcr}}\,}
\fi
}

\AtBeginDocument{%
  \@ifpackageloaded{endfloat}%
   {\renewcommand\efloat@iwrite[1]{\immediate\expandafter\protected@write\csname efloat@post#1\endcsname{}}}{\newif\ifefloat@tables}%
}%

\def\BreakURLText#1{\@tfor\brk@tempa:=#1\do{\brk@tempa\hskip0pt}}
\let\lt=<
\let\gt=>
\def\processVert{\ifmmode|\else\textbar\fi}

\@ifundefined{subparagraph}{
\def\subparagraph{\@startsection{paragraph}{5}{2\parindent}{0ex plus 0.1ex minus 0.1ex}%
{0ex}{\normalfont\small\itshape}}%
}{}

\newcommand\role[1]{\unskip}
\newcommand\aucollab[1]{\unskip}
  
\@ifundefined{tsGraphicsScaleX}{\gdef\tsGraphicsScaleX{1}}{}
\@ifundefined{tsGraphicsScaleY}{\gdef\tsGraphicsScaleY{.9}}{}
\def\checkGraphicsWidth{\ifdim\Gin@nat@width>\linewidth
	\tsGraphicsScaleX\linewidth\else\Gin@nat@width\fi}

\def\checkGraphicsHeight{\ifdim\Gin@nat@height>.9\textheight
	\tsGraphicsScaleY\textheight\else\Gin@nat@height\fi}

\def\fixFloatSize#1{}%\@ifundefined{processdelayedfloats}{\setbox0=\hbox{\includegraphics{#1}}\ifnum\wd0<\columnwidth\relax\renewenvironment{figure*}{\begin{figure}}{\end{figure}}\fi}{}}
\let\ts@includegraphics\includegraphics

\def\inlinegraphic[#1]#2{{\edef\@tempa{#1}\edef\baseline@shift{\ifx\@tempa\@empty0\else#1\fi}\edef\tempZ{\the\numexpr(\numexpr(\baseline@shift*\f@size/100))}\protect\raisebox{\tempZ pt}{\ts@includegraphics{#2}}}}

\AtBeginDocument{\def\includegraphics{\@ifnextchar[{\ts@includegraphics}{\ts@includegraphics[width=\checkGraphicsWidth,height=\checkGraphicsHeight,keepaspectratio]}}}

\DeclareMathAlphabet{\mathpzc}{OT1}{pzc}{m}{it}

\def\URL#1#2{\@ifundefined{href}{#2}{\href{#1}{#2}}}

\def\UrlOrds{\do\*\do\-\do\~\do\'\do\"\do\-}%
\g@addto@macro{\UrlBreaks}{\UrlOrds}

\edef\fntEncoding{\f@encoding}

\makeatother

\newif\ifmultipleabstract\multipleabstractfalse%
\usepackage[section]{placeins}

\usepackage{float}

\begin{document}

\begin{frontmatter}

\title{
    CVEH: A Dynamic Framework To Profile Vehicle Movements To Mitigate Hit And Run Cases Using Crowdsourcing    
}
    
\author[]{Attiq ur Rehman}
\ead{attiq.qadeer@gmail.com}
\author[]{Asad Waqar Malik}
\ead{asad.malik@seecs.edu.pk}
\author[]{Anis ur Rahman}
\author[]{Sohail Iqbal}
\ead{sohail.iqbal@seecs.edu.pk }
\author[]{Ghalib Ahmed Tahir}
\ead{12mscsgtahir@seecs.edu.pk}
    
\address{
    National University Of Science And Technology}

\begin{abstract}
In developed countries like the USA, Germany, and the UK, the security forces used highly sophisticated equipment, fast vehicles, drones, and helicopters to catch offenders' vehicles. Whereas, in developing countries with limited resources such schemes cannot be utilized due to management cost and other constraints. In this paper, we proposed a framework called CVEH that enables developing countries to profile the offender vehicle movements through crowdsourcing technique and act as an early warning system to the law forcing agencies. It also engages citizens to play their role in improving security conditions. The proposed CVEH framework allows Vehicle-to-Infrastructure (V2I) communication to monitor the movement of the offender's vehicle and shared its information with the Command and Control (CC) centre. The CC centre projects the path and engages nearly located law enforcement agencies. CVEH is developed and evaluated on android smartphones. Simulations conducted for this study exhibit the effectiveness of our framework.

\end{abstract}
\end{frontmatter}
    
\section{Introduction}
Crowdsourcing is an outsourcing model within which people or organizations obtain contributions from the users (maybe over the internet) to get specific ideas and services. This kind of outsourcing to attain higher results by dividing the tasks between the users had already gained success even before the digital era. However, subcontracting in crowdsourcing is discriminated against by the statement that in crowdsourcing, an undetermined group of people do the jobs rather than a specifically tasked cluster as crowdsourcing uses a mix of top-down and bottom-up approaches. The blessings of crowdsourcing embrace better prices, agility, fineness, tractability, measurability, or variety. Figure 1 [28] depicts the design architecture of crowdsourcing.

\bgroup
\fixFloatSize{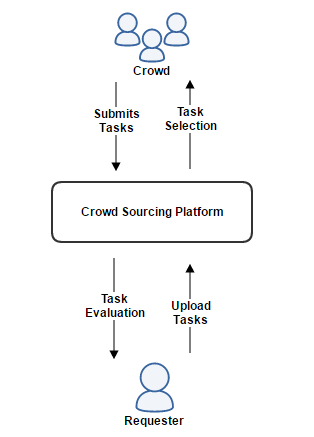}
\begin{figure*}[!htbp]
\centering \makeatletter\IfFileExists{images/crowdsourcingarchitecture.png}{\includegraphics{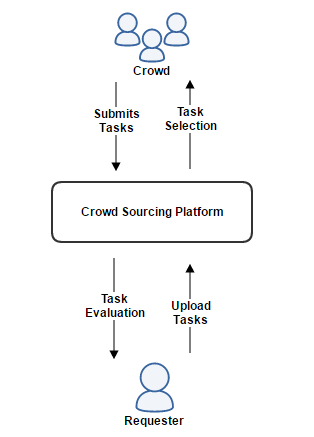}}{}
\makeatother 
\caption{{Crowdsourcing Architecture}}
\label{f-ed4ea42f64d6}
\end{figure*}
\egroup
The target of our research is to use crowdsourcing for the advantage of the community. In underdeveloped nations, it's impractical to install a gigantic foundation for the monitoring and tracking systems in this manner, CVEH is one of the conceivable ways to deal with defeat framework issue and help in accomplishing social manageability.

\subsection{Motivation }In the post 9/11 world, the market for camera-based video surveillance products has expanded manifold with its primary use as an area monitoring system by public safety and security organizations \unskip~\cite{931999:20758532}. Though such systems are cheap and ubiquitous, there are limitations and underlying challenges in deploying the practical system. These challenges involve the cost of wiring, low power hardware and the development of detailed management tools \unskip~\cite{931999:20766508}. Once installed, the human resource required to monitor many real-time feeds simultaneously is expensive.  In many cases, this becomes ineffective due to the limits of human surveillance \unskip~\cite{931999:20766560}. Here automated video surveillance helps by automatically analyze a scene and generate alarms for the manual supervisor to take action. No doubt, such systems based on video analytics are the future, but very few are robust enough to work in variations and unfavourable conditions \unskip~\cite{931999:20766587}. Moreover, existing systems such as VSAM (Video Surveillance and Monitoring) \unskip~\cite{931999:20766625} faces several challenges in their underlying components, which impacts the overall efficiency.

\subsection{Problem Statement}With the urban growth in Pakistan, there is a substantial increase in the number of crimes involving vehicles, ranging from road violations to serious crimes. In practice, there are many sensor-based solutions primarily used for road traffic estimation and congestion detection. Moreover, some of these systems identify minor traffic violations, e.g. an automated parking metering system issuing a parking ticket on an expired meter but not used by law enforcement agencies. \mbox{}\protect\newline For crime detection, the use of automated vehicle surveillance to identify and track suspicious vehicles is becoming more and more important. Though wide-area camera-based video surveillance can help perform this task with increasing workload and other overhead expenses, manual visual tracking is impossible. Mainly such video surveillance is treated as a post-incident forensic analysis tool, with investigators going through hordes of recorded data, which is a serious problem impacting the effectiveness of doing real-time surveillance. An automated system capable of triggering alarms in this situation can be useful; however, depending on the nature of the scene, the system may generate many false alarms that need to be assessed by a human supervisor\unskip~\cite{931999:20766669}\unskip~\cite{931999:20766692}\unskip~\cite{931999:20766693}\unskip~\cite{931999:20766703}. \mbox{}\protect\newline For crime detection and deterrence, as a replacement for the old system of using a police gazette with rewards, the distributed human intelligence model can work more effectively. For instance, after the Boston Marathon bombing, an open call to share the content of the incident helped to apprehend the culprits within few days \unskip~\cite{931999:20766706}. Similarly, they employed the same approach after the London riots in 2011 \unskip~\cite{931999:20766709}. Though there are controversies, from false accusations to rumours, a moderated platform can be used to put citizens immediately to work by sifting through video recordings with the investigators directing the analysis by shutting down misinformation \unskip~\cite{931999:20766710}. Moreover, law enforcement and the public can use collected data to identify areas with high crime rates.

\subsection{Contribution Of The Study }In this paper, our contribution is on vehicle identification and reporting mechanisms through crowdsourcing. We proposed CVEH to profile the movement of offender vehicles. The command and control centre utilizes its limited resources efficiently by identifying the offender direction and project its path through extrapolation. Our approach provides an affordable and easy to maintain surveillance system to the developing countries which cannot afford complex and advanced surveillance systems that use advanced and expensive technologies. Our proposed methodology is capable of providing surveillance capabilities to small streets and neighbourhoods where the installation of dedicated surveillance systems is not feasible. It also helps create a sense of ownership and responsibility among the public and therefore increases the coherence between law enforcing agencies and the general public.

\subsection{Outline}The paper is outlined as follows. In section 3 we have reviewed the literature and related work done in this field. Section 4 describes the framework and proposed design while conclusion and future work is described in section 5. 
    
\section{Related Work}
Smart cities merge information and communication technology and the Internet of Things securely. The concept is to attach all the city assets in a grid for monitoring, controlling and utilizing efficiently. In the recent decade, there has been significant work in various domains of smart cities. \mbox{}\protect\newline Ibrar Y. et al. \unskip~\cite{931999:20766711} discussed the communication and connectivity problem in smart cities. The backbone of smart cities is the communication channel, and in the presence of millions of devices, communication technology cannot provide flawless connectivity. Moreover, the authors have also presented the case study of Stratford, Singapore, etc., as a proof of concept.

M. Victoria Moreno et al. \unskip~\cite{931999:20766749} discussed the role of big data in smart cities. The authors proposed a four-layered architecture for smart cities to handle data coming from static and dynamic sensors to utilize energy efficiently. Similarly, Z. Ali et al. \unskip~\cite{931999:20766750} proposed an automatic voice complaint system to monitor the resident of the smart home. The concept is detecting emergency, especially in homes where no one is available to help aged people. The proposed detection scheme employs the theory of linear prediction analysis. 

 Smart transport systems are playing a vital role in the last decade. Significant progress is done in terms of route optimization, congestion avoidance. Z. Bhutto et. al \unskip~\cite{931999:20766850} used smartphones in place of CCTV cameras to broadcast live videos to help reduce street crimes. It enables users to stream live videos of street crimes using3G/4G LTE or a Wi-Fi connection. The live streaming of the video feed into the police servers. Thus, the people participated in reporting the crimes to the law enforcement agencies and hence help in capturing the offenders. The Real-Time Streaming Protocol (RTSP) is used to transmit data in packets to a media server.

S. Mujtaba et al. \unskip~\cite{931999:20766851} proposed a framework to report street crimes using mobile phones. The initiator can generate a crime report by taking a picture and sending it to the server with a crime tag. They also sent the Geolocation of the reports to the server. GPS technology built-in in most mobile phones tagged the location. It uses 3G/4G or Wi-Fi technology to transmit data to and from the server. After the report generation, all the registered users can see the crime location on the map in the app and take precautionary measures.

Divya Lal et al. \unskip~\cite{931999:20766852} proposed an app that provides the option to report crimes by using multiple methods. As a first step to use the app, users need to register themselves. To report a crime, the user can record a voice message, capture a video or image and initiate the report. This information is transmitted along with the location to the police station to keep track of users. The application periodically sends the GPS location of the user to the server. Similarly, in \unskip~\cite{931999:20766854}, Alexey Medvedev et al. proposed a framework called CityWatchers, using vehicle-mounted surveillance cameras to generate alerts in case of an accident, traffic jams, etc. Drivers initiate the alerts, and users receive the incentives based on their participation. \mbox{}\protect\newline W. A. Agangiba et al. \unskip~\cite{931999:20766855}  implemented the idea of using the computational capabilities of smartphones to make them useful for society for crime detection and reporting. The proposed framework helps in the detection, reporting and eventually tracking the offenders using smartphones. In \unskip~\cite{931999:20766856}, M. Eunus et al. introduced SafeStreet, a smartphone application that used crowdsourcing and GPS location service. This application helps women to report sexual harassment cases in public places. It enables women to capture and share their experiences anonymously. Users can also find a path to a safe location. It stores All the reported cases on the server with the location and time of the incident. Based on analytics, it can also recommend safe travel time.

Wisam A. F. et al. \unskip~\cite{931999:20766858} proposed a methodology to capture digits from dissimilar license plates. The authors proposed model is based on fuzzy rules to identify the fragmented characters and numerical digits from the same input sets. Thus, the technique is to identify license plates that are not following the standard template. Using the hybrid method of the Fuzzy template matching scheme, the authors attained 90.4\% accuracy in an outdoor environment. 

Similarly, OpenALPR is an open-source Automatic License Plate Recognition library available in different languages. Sandro Machado et al. \unskip~\cite{931999:20766860} ported the OpenALPR library to Android smartphones. The library analyzes images and video streams to identify license plates. The output is the text representation of any license plate characters. Integration of the automatic license plate recognition (ALPR) system and crime reporting system on smartphones is the new concept in reporting and tracking crime incidents using smartphones. \mbox{}\protect\newline  \mbox{}\protect\newline In \unskip~\cite{931999:20766856}, M. E. Ali et al. have proposed SafeStreet smartphone-based application that uses crowdsourcing and GPS location service. Women can report sexual harassment cases through the application.  It provides the ability to capture and share their experiences anonymously. It also helps users to find a safe path. They stored all the reported harassment on the server with the location and time of the incident for further analysis. \mbox{}\protect\newline Alessandro Amoroso et al. \unskip~\cite{931999:20766861}  present a model in which people establish an ad hoc network using smartphones. It implements an optimal and lightweight P2P communication system using GPS and Wi-Fi functionalities of smartphones. \mbox{}\protect\newline Matthias Pfriem et al. \unskip~\cite{931999:20766862}  have present the approach to use smartphones as a multi-sensor platform in a field operational test for naturalistic driving study. Modern smartphones offer the sensor capability to detect speed and position as well as accelerations. These sensors of smartphones measure the mobility of vehicles. Hence, they do the task by crowdsourcing without using dedicated data loggers. \mbox{}\protect\newline Eduardo Romero et al. \unskip~\cite{931999:20766863} presents a vision-based vehicle detection and tracking. The approach employs smartphone cameras as an alternative to other expensive and complex sensors such as LiDAR or other range-based methods. A multi-scale proposal and simple geometry consideration of the roads based on the vanishing point are combined to overcome the computational constraints. They tested their algorithm on a publicly available road data set, thus demonstrating its real applicability to autonomous driving. \mbox{}\protect\newline Akhil M. et al. \unskip~\cite{7447636} propose an inexpensive and effective solution for traffic signals. The proposed scheme employs vehicle detectors (VD's) near intersections. When a vehicle approaches the intersection, it sequentially passes the VD's. VD's then transmit their unique IDs to the smartphone. The smartphones then send the  IDs and relevant kinematic data to a base station installed at the intersections. Thus, the base station can accurately compute the expected times of the vehicles at the intersection. In the event of a potential collision, the base station can transmit a warning to the smartphones of the concerned drivers. 
    
\section{CVEH: A dynamic framework to profile vehicle movements to mitigate hit and chase cases using crowdsourcing}
CVEH allows users to participate, in capturing the offenders, generate alerts and play its role in law and order implementation. In developed countries, high-speed vehicles in coordination with helicopters, quadcopters and surveillance cameras placed at various locations used to capture fast-moving offenders. Whereas, in developing countries, limited resources are available to maintain the law and order situation. These countries cannot afford high-speed vehicles due to their cost and maintenance. Moreover, it is also unwieldy to install cameras at every intersection. Thus, CVEH provides a platform to maintain law and order under such conditions through the crowdsourcing paradigm. Without proper surveillance, in a typical hit-and-run case, vehicles can disappear in urban areas. Therefore, our proposed frameworks help in identifying the offender's location and project its movement through crowdsourcing. This approach gives an early warning to law enforcement agencies to take preemptive action to capture the offender. Figure 2 shows the high-level architecture of the CVEH framework. It consists of two main modules user module and the command control centre, which are explained as follows.

\bgroup
\fixFloatSize{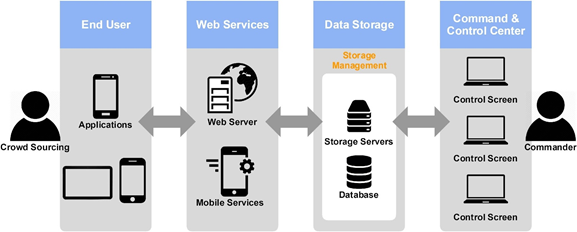}
\begin{figure*}[!htbp]
\centering \makeatletter\IfFileExists{images/image.png}{\includegraphics{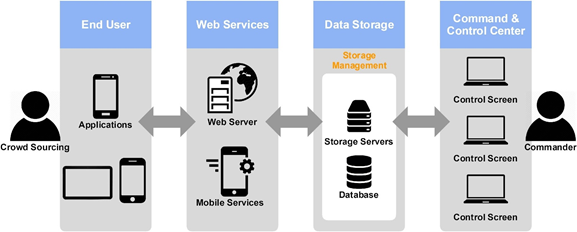}}{}
\makeatother 
\caption{{CVEH Architecture}}
\label{f-bcdf5671eed7}
\end{figure*}
\egroup

\subsection{Proposed Design}The proposed framework facilitates end-users to launch complaints through the web or mobile applications. The timeline diagram in Figure 3 shows the flow of events between users and the command control module. \mbox{}\protect\newline CVEH framework facilitates developing countries to arrest offenders through crowdsourcing. The system consists of two major modules, the user module and Command Control (CC) Center. The user module is the main part that executes on the handheld device. As proof of concept, they designed the CVEH user module for the Android operating system. As a first step, the user needs to download and register for the CVEH client application. The registration process authenticates the user through its national security number. This registration process is mandatory to avoid any fake/fabricated user input. Thus, the user can use CVEH on any handheld device issued to its name. Similarly, the CVEH application can also work on Android-based vehicle-mounted surveillance systems. We also assume that smartphones are mounted on the vehicles' dashboards and can act as a surveillance system. Moreover, the pedestrian can use CVEH in manual surveillance mode.

\bgroup
\fixFloatSize{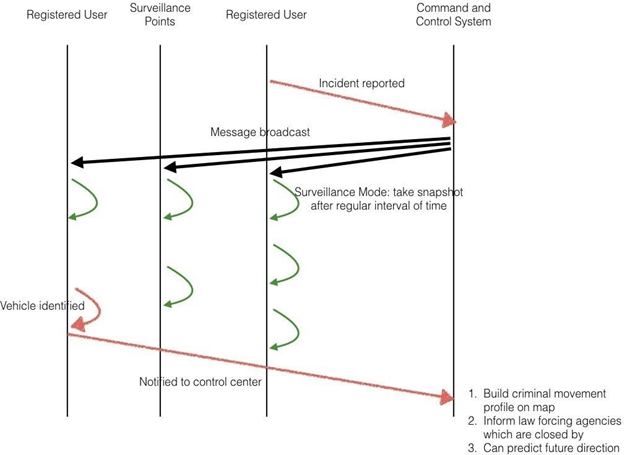}
\begin{figure*}[!htbp]
\centering \makeatletter\IfFileExists{images/0e60edfe-dab0-46ae-be75-e319b9bcdb3a.png}{\includegraphics{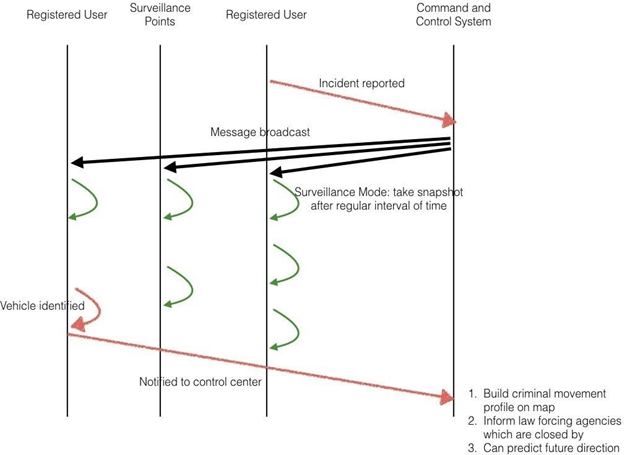}}{}
\makeatother 
\caption{{CVEH application flow}}
\label{f-27a090451e8a}
\end{figure*}
\egroup
After successful authentication, the user can play its role in sensing the environment and help to implement the law and order situation. In case of any incident like theft, mobile snatching, etc. the offenders use vehicles to move out of the crime location. The CVEH user can capture the crime scene through a snapshot of the offender and its assisted vehicle. The user can also enter the required information manually. In pedestrian mode, the user can take a snapshot of a vehicle and CVEH app imaging module extracts the vehicle registration number from the image using Optical Character Recognition (OCR) module. Figure 4 depicts the process.

\bgroup
\fixFloatSize{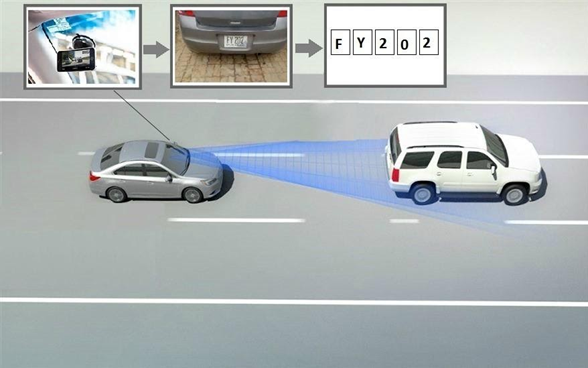}
\begin{figure*}[!htbp]
\centering \makeatletter\IfFileExists{images/snapshotmodule.png}{\includegraphics{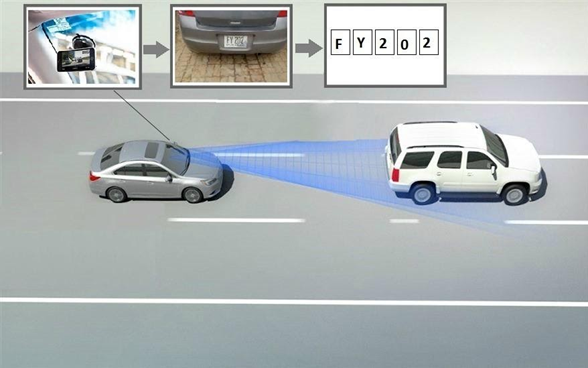}}{}
\makeatother 
\caption{{Snapshot Module }}
\label{f-1b5e6678b3dd}
\end{figure*}
\egroup
The user can enter other details of the vehicle such as colour, make, model, etc. It sends all the information to the command and control centre. Moreover, the command and control centre is linked with the online vehicle repository system. The shared information also contains the location of the incident extracted through GPS and the time of the incident. After receiving the complaint, it generates an alert, which is broadcasted to all the users present in the near vicinity i.e. within 10 km (configurable). When there is no response within 15 minutes of the alert, the systems broadcast the message to all the registered users within the city; this is to identify the initial location of the suspect's vehicle. Figure 5 illustrates the registration, reporting and broadcasting functionality.

\bgroup
\fixFloatSize{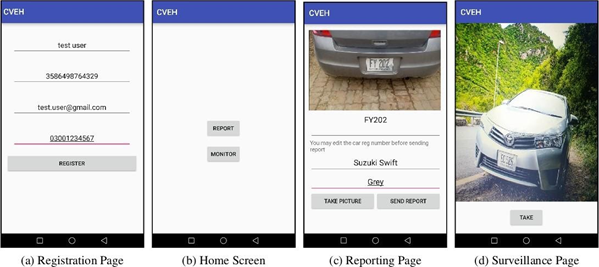}
\begin{figure*}[!htbp]
\centering \makeatletter\IfFileExists{images/8a0112f7-cde7-4355-aa0a-2ded386be8d4.png}{\includegraphics{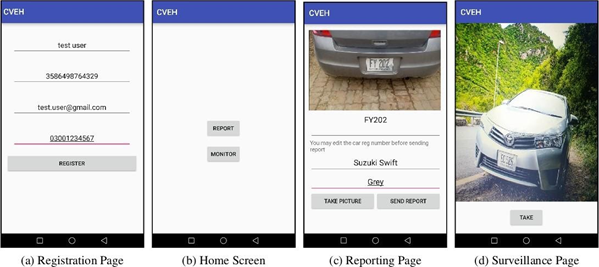}}{}
\makeatother 
\caption{{Images of Different Screens of Our Prototype Application}}
\label{f-bc6819973af7}
\end{figure*}
\egroup
On receiving the alert from the command and control centre, the CVEH application configured to work in a surveillance mode mounted on the vehicle's dashboard starts taking the snapshot after a regular interval of time. In this mode, the CVEH surveillance module takes a snapshot of the surroundings, extract the vehicle number and matches the received alert. Notification pop-up to inform the users about the alert, in CVEH handheld device configuration mode. The user can contribute by manually taking a snapshot of its surroundings. On successful identifying offenders' vehicle, it sends the information to the control room. The control room profile the movement of the offender vehicle based on the crowd input and alert the law enforcing agencies that are located at an approachable distance from the offender. It calculates the projected path at run time through extrapolation. The command and control centre are equipped with computing machines where it displays the user's input in the form of offender movement profiling. The control centre also includes a database where it stores all the information for further analysis. Inside the command and control centre, the Law Enforcing Agencies (LEA's) can also see the location of their registered users, reported incidents on the map and can also directly communicate with registered users, as shown in Figure 6. Moreover, the CVEH application provides easy integration with surveillance cameras installed at various intersections; thus, we can utilize the installed camera similarly as discussed above.

\bgroup
\fixFloatSize{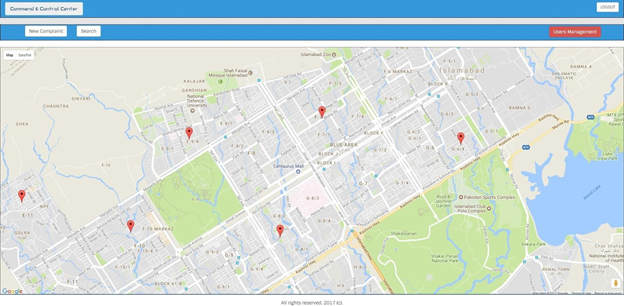}
\begin{figure*}[!htbp]
\centering \makeatletter\IfFileExists{images/6f793d93-a842-4bed-bfeb-9a34a2fcd6d0.png}{\includegraphics{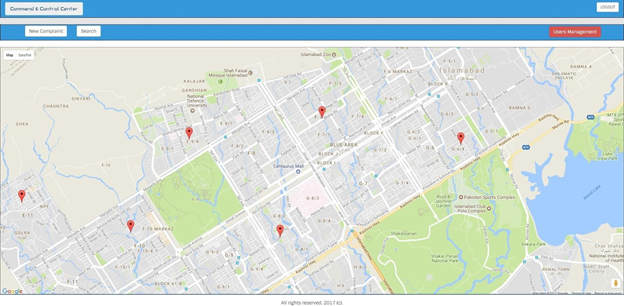}}{}
\makeatother 
\caption{{Command and Control Center, Markers are Crowd sourced Points}}
\label{f-bf28c8c1a6a9}
\end{figure*}
\egroup
The OPEN ALPR (Android) library extract the vehicle information from the snapshot. Typically, there can be multiple vehicles in the snapshot. Therefore, correct extraction of all the vehicle information is deemed necessary. Moreover, to multicast message/alerts, Firebase Cloud Messaging (FCM) is used. The OpenALPR is an Open source Automatic License Plate Recognition library written in C++ with bindings available in other languages. The library analyzes images and video streams to identify license plates, and the output is the text representation of any license plate characters. The Fire-based Cloud Messaging (FCM) is a cross-platform reliable messaging solution. The FCM facilitates generating notification messages to engage and retain users.  It can support message payload up to 4KB. \mbox{}\protect\newline

\subsection{\textbf{Case study: ICT Surveillance System}}As proof of concept, we have studied the surveillance system installed in Islamabad, the capital of Pakistan. In Islamabad city, there is a lack of surveillance infrastructure, especially in the inner-city area. Only partial infrastructure is available at entries and exits. We used the standard vehicle number plates, as shown in Figure 7. Our approach allows CVEH to easily identify vehicles without using complex machine learning algorithms executing at handheld devices. Moreover, it installs few surveillance cameras at various locations that can become part of the CVEH grid.

\bgroup
\fixFloatSize{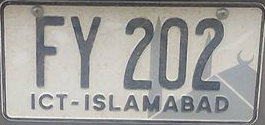}
\begin{figure*}[!htbp]
\centering \makeatletter\IfFileExists{images/c184fa6f-7248-41a2-b201-4a78e753a1c8.png}{\includegraphics{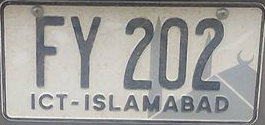}}{}
\makeatother 
\caption{{Standard Vehicle Number Plate for Islamabad }}
\label{f-f1bb24a8439b}
\end{figure*}
\egroup

\subsubsection{Prototype evaluation}To evaluate the feasibility of the proposed CVEH framework, we have analyzed the system based on its resource consumption. The prototype application implements the core functionality of the system. We have developed a client (pedestrian and vehicle dashcam surveillance) mode and command and control centre. Our application uses the OpenALPR library to provide  OCR functionality. The CVEH module at the client end automatically detects the license plate from the image and extracts the vehicle registration number from the license plate. The execution of CVEH at the client end requires memory, processing power, and a data package. In the benchmark study, we kept the client application running for consecutive 23 days and continuously measure the resource usage of the proposed framework. Figure 8 (a) shows the time taken in generating a request and receiving a response from the control centre (we reported only on 100 requests). In all the communication, the average time is around 30 secs (we obtain some of these readings during peak hours). We also observed the memory, battery and CPU consumption during the experiments. Figure 8 (d) shows the CPU usage is a little high at places where multiple vehicles detected in a single snapshot. We reported the average CPU consumption of the entire day at discrete intervals. Moreover, the memory and battery usage are less than 4 {\textendash} 8 per cent respectively. To reduce the memory consumption, we have stored limited information at the CVEH client end. Most of the data is stored in the form of the tree structure to reduce the query time. Similarly, the battery is directly related to resource consumption which keeps all the resource usage below the threshold value as shown in Figure 8.

\bgroup
\fixFloatSize{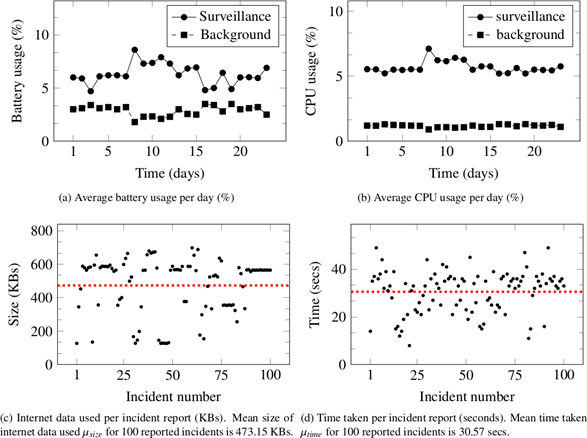}
\begin{figure*}[!htbp]
\centering \makeatletter\IfFileExists{images/resourceusage.png}{\includegraphics{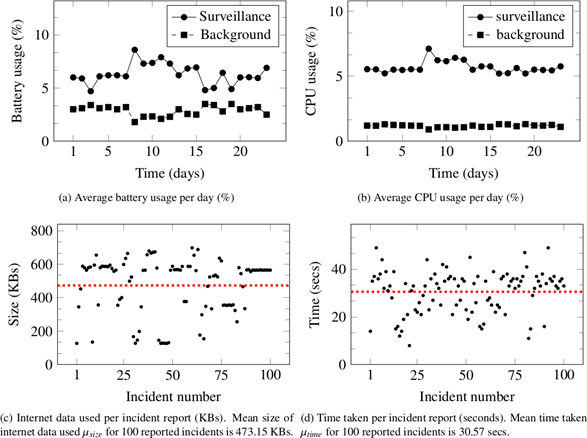}}{}
\makeatother 
\caption{{Resource usage statistics}}
\label{f-1c8ae2f5edcb}
\end{figure*}
\egroup
To extract the vehicle registration number from the image taken, we have used a variant of OPEN ALPR for Android to tune it for the new Islamabad Capital Territory (ICT) license format and used FCM to broadcast the message to the users. Moreover, we evaluated the proposed system through simulations in Matlab, where different crowdsource points are placed on a google map to identify the offenders' vehicle. In the simulation, we increased the source points from 100 to 1000, and all the source points are movable. \mbox{}\protect\newline The results show that with an increase in the crowdsource points, the probability of identifying vehicle also increases, as shown in Figure 9. It means more volunteers can significantly improve the performance of the proposed system. Hence, with the increase in the number of crowdsource points, the system efficiency is increased. It can do the surveillance with the minimal cost of infrastructure, which is very helpful for developing nations.

\bgroup
\fixFloatSize{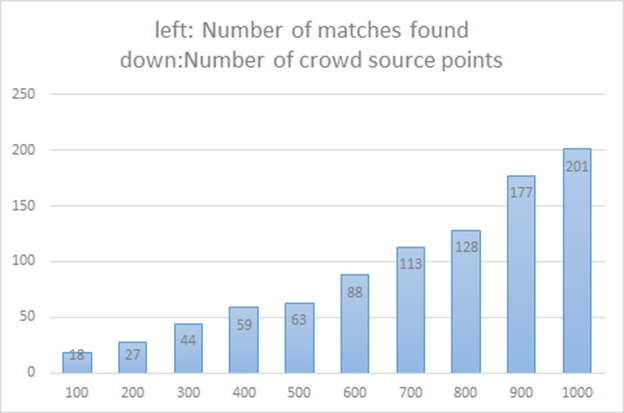}
\begin{figure*}[!htbp]
\centering \makeatletter\IfFileExists{images/figure9.png}{\includegraphics{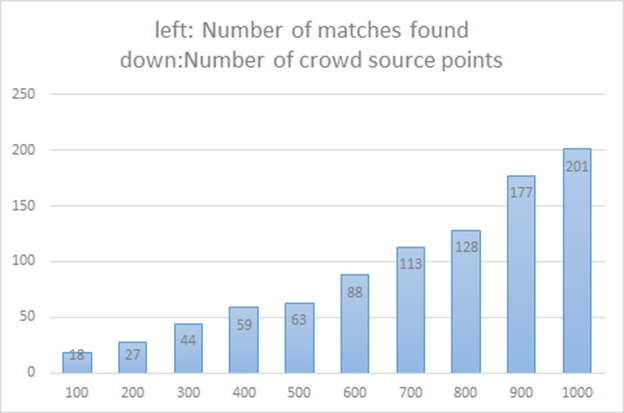}}{}
\makeatother 
\caption{{}}
\label{f-4f35fb6dbf61}
\end{figure*}
\egroup

\section{Conclusion}
In this paper, we proposed the CVEH framework to help developing countries for the sustainability of society. Our proposed framework is robust and scalable. To evaluate our architecture, we have implemented a prototype that works on Android smartphones and benchmarks the resource's consumption. Moreover, we have also simulated the effect of crowdsourced points on vehicle detection, which indicate that the efficiency and accuracy of the system are directly proportional to the number of crowdsource points. The objective of CVEH is to utilize volunteers for the benefit of society. In developing countries, it is not feasible to deploy huge infrastructure for surveillance. Therefore, CVEH is one of the possible approaches to overcome infrastructure problems and help in achieving social sustainability.

\textbf{Sustainable Development Goals (SDG's): }The United Nations has declared a set of 17 goals as the sustainable Development Goals (SDG's). These goals and targets consisted of a large collection of sustainable development matter, e.g. eliminating poverty and famine, fighting climate change, look after the oceans and forests, etc. Our application facilitates the users to report the incidents in a short amount of time. As the law enforcement institutes can provide swift justice and it helps a country to achieve the sustainable development goal ``Peace, Justice and Strong Institutions'' by making the environment safer for the people.
    
\section{Future Work}
In the future extension of this work, we will improve the algorithm accuracy, for detection in low visibility environments like nighttime with low lighting or during thunderstorms where there are hindrances in clear vision. The extension of the application on iOS smartphones is in the pipeline. Other proposed extensions include dashcam support, integration with advanced cameras available at important public places and integration with national citizens database to regularize the registration process, provide a safety layer from prank/fake reports and make identification and surveillance easier for the law enforcement agencies.

\bibliographystyle{elsarticle-num}

\bibliography{\jobname}

\end{document}